\documentstyle[12pt,aasms4,graphics]{article}
\newcommand{\etal}{et~al.}

\begin{document}
\oddsidemargin=0mm

\title{
A {\it Hubble Space Telescope} Snapshot Survey of Nearby Supernovae
\footnote{ Partially based on observations with the NASA/ESA {\it Hubble Space
Telescope}, obtained at the Space Telescope Science Institute, which is
operated by the Association of Universities for Research in Astronomy
(AURA), Inc., under NASA contract NAS 5-26555.}}

\author{Weidong Li\altaffilmark{2}, Alexei V. Filippenko\altaffilmark{2}, 
Schuyler D. Van Dyk\altaffilmark{3}, 
Jingyao Hu\altaffilmark{4}, Yulei Qiu\altaffilmark{4},
Maryam Modjaz\altaffilmark{2,5}, and Douglas C. Leonard\altaffilmark{2,6}\\
Email: (wli, alex)@astro.berkeley.edu}

\altaffiltext{2}{Department of Astronomy, University of California, Berkeley,
CA 94720-3411.}

\altaffiltext{3}{Infrared Processing and Analysis Center, 100-22 Caltech,
Pasadena, CA 91125.}

\altaffiltext{4}{Beijing Astronomical Observatory, Chinese Academy of Science,
Beijing 10012, China.}

\altaffiltext{5}{Present address: Center for Astrophysics, 60 Garden St.,
Cambridge, MA 02138.}

\altaffiltext{6}{Present address: Department of Astronomy, University of
Massachusetts, Amherst, MA 01003-9305.}

\begin{abstract}

We present photometry of 12 recent supernovae (SNe) recovered in a {\it Hubble
Space Telescope} Snapshot program, and tie the measurements to earlier
ground-based observations, in order to study the late-time evolution of the
SNe. Many of the ground-based measurements are previously unpublished, and were
made primarily with a robotic telescope, the Katzman Automatic Imaging
Telescope.  Evidence for circumstellar interaction is common among the
core-collapse SNe.  Late-time decline rates for Type IIn SNe are found to span
a wide range, perhaps due to differences in circumstellar interaction. An
extreme case, SN IIn 1995N, declined by only 1.2 mag in $V$ over about 4 years
following discovery. Template images of some SNe must therefore be obtained
many years after the explosion, if contamination from the SN itself is to be
minimized. Evidence is found against a previous hypothesis that the Type IIn SN
1997bs was actually a superoutburst of a luminous blue variable star. The
peculiar SN Ic 1997ef, a ``hypernova," declined very slowly at late times. The
decline rate of the SN Ia 2000cx decreased at late times, but this is unlikely
to have been caused by a light echo.

\end{abstract}


\section{INTRODUCTION}

Supernovae (SNe) represent the final, explosive stage in the evolution of
certain varieties of stars (see, e.g., Woosley \& Weaver 1986; Arnett
\etal~1989; Wheeler \& Harkness 1990; Filippenko 1997, for reviews of SN types
and explosion mechanisms). They synthesize and expel heavy elements, heat the
interstellar medium, trigger vigorous bursts of star formation, create neutron
stars and sometimes black holes, and produce energetic cosmic rays. Type Ia SNe
(SNe Ia), among the most luminous of SNe, are exceedingly useful cosmological
tools and have been used to study the expansion history of the Universe.  These
studies (Riess \etal~1998, 2001; Perlmutter \etal~1999; see Filippenko 2001 for
a recent summary) reveal the surprising result that the expansion of the
Universe is currently accelerating, perhaps due to a nonzero cosmological
constant. SNe are clearly among the most interesting and important constituents
of the Universe and should be vigorously studied.

\subsection{Discovery and Monitoring of Nearby Supernovae}

Detailed spectral and photometric observations of SNe can be used to
study the properties of SNe, compare SNe at different redshifts, and
gain insights into the evolutionary paths that lead to these energetic
explosions.  However, until recently, most relatively nearby SNe were
found either sporadically, in images taken for other purposes, or at a
considerable time after the explosion. Moreover, systematic follow-up
observations were scarce for all but a minor fraction of the
SNe. During the past few years, however, the situation has changed
dramatically, as robotic (or nearly robotic) telescopes have been
dedicated to the search for and follow-up of SNe (Filippenko
\etal~2001).  The two outstanding examples are the Beijing
Astronomical Observatory Supernova Search (BAOSS; Li \etal~1996) with
a 0.6-m telescope, and the Lick Observatory Supernova Search (LOSS;
Treffers \etal~1997; Li \etal~2000; Filippenko \etal~2001) with the
0.8-m Katzman Automatic Imaging Telescope (KAIT).

KAIT reaches a limit of $\sim$ 19 mag (4$\sigma$) in 25-s unfiltered, unguided
exposures, while 5-min guided exposures yield $R \approx $ 20 mag. 2500$-$7000
galaxies (most with $cz \leq$ 6000 km s$^{-1}$) are surveyed during any
particular season, with a cycle time of $\sim$ 3$-$10 days. The search software
automatically subtracts new images from old ones and identifies SN candidates
which are subsequently examined by undergraduate research assistants.

LOSS discovered its first SN in 1997 (SN 1997bs; Treffers \etal~1997; Van Dyk
\etal~2000), then 20 SNe in 1998, 40 in 1999, 36 in 2000, and 68 in
2001. Together, LOSS and BAOSS have found a total of over 180 SNe in the past 5
years. Recently, LOSS teamed up with Michael Schwartz in Arizona, forming the
Lick Observatory and Tenagra Observatory Supernova Search (LOTOSS), allowing
more SNe to be discovered and followed.

Since most of the SNe found during these systematic searches are discovered
very early in their development, and have a large amount of follow-up time
devoted to them, they are among the world's best-studied SNe (see, e.g., Li
\etal~2000, 2001a; Leonard \etal~2002a,b). Li, Filippenko, \& Riess (2001) and
Li \etal~(2001b) also show that these surveys are least affected by
observational biases, and provide the most accurate luminosity function for
SNe. The rate of intrinsically peculiar SNe~Ia, for example, is found to be
unexpectedly high.

\subsection{The Environment and Late-Time Detection of Supernovae}

The explosion of a SN leaves few traces of the star that underwent the
catastrophic event. An important clue to the nature of the progenitor is its
environment (e.g., Boffi 1999). Unfortunately, most studies of the sites of SNe
have been hampered by the limited spatial resolution of ground-based
observations (e.g., van den Bergh 1988; Panagia \& Laidler 1991; Boffi, Sparks,
\& Macchetto 1999). This problem can be partially overcome by using {\it Hubble
Space Telescope (HST)} archival images (e.g., Van Dyk \etal~1999a,b, 2000).
However, although the chance that the site of any particular SN has been imaged
by {\it HST} during other programs is rapidly growing, it is still relatively
low. Also, some of the positions of older SNe (prior to 1990) were uncertain,
compromising studies of their local environment.

To remedy this, an {\it HST} Snapshot survey program (GO-8602) of recent,
nearby SNe was conducted starting in Cycle 9, requesting WFPC2 images of the
sites of 23 LOSS and BAOSS SNe\footnote{Some of these 23 SNe were first
discovered by other groups, but subsequently found in the course of LOSS and
BAOSS.} with $cz \leq $6000 km s$^{-1}$.  These SNe provide the best spatial
resolution and have accurate positions (often better than $\pm 0\farcs 5$). In
addition, they were generally discovered early in their development, and had
extensive ground-based follow-up studies. 45 out of the 70 ($\sim 45\%$)
requested observations have been made, with 20 out of the 23 SNe ($\sim 90\%$)
observed at least once and 13 ($\sim 56\%$) of the SNe having images in more
than one filter; thus, the chances of imaging the sites of specific SNe are
much higher than by using the archive alone. Moreover, the data are all
obtained in a uniform way using the same set of filters.

One advantage of concentrating the Snapshot survey on very recent, nearby,
well-studied SNe is that many of the SNe themselves are still visible in the
$HST$ images, providing late-time photometry superior to that achieved from the
ground; at such late times, the SNe are so faint that their ground-based
photometry is contaminated by neighboring stars within the seeing
disk. Late-time photometry, especially through more than one filter, provides
not only useful information on the underlying physics for the lingering light,
such as radioactive decay of long-lived isotopes, interaction with
circumstellar matter, and light echoes, but also the relative brightness of SNe
still present in ``template" images taken at various epochs compared to their
maximum light. To obtain proper photometry of a SN, which often occurs in a
complicated background (e.g., spiral arms or H~II regions), observers are
required to take template images of the host galaxy a year or two after the
discovery, and then do image subtraction. It is often assumed that the light of
the SN is essentially negligible in these template images, but as we later
discuss in this paper, some SNe (especially those of Type II) are quite
long-lived and may contaminate the template images.

Detailed analysis of the SN environments is still underway, and the results
will be discussed elsewhere.  In this paper, we report on SNe recovered in the
Snapshot images. Section 2 contains a description of the observations and
analysis of the photometry, while \S~3 presents the light curves of all
recovered SNe, details of individual SNe, and comparison of late-time
light-curve shapes and decline rates. Our conclusions are summarized in \S~4.

\section{OBSERVATIONS AND REDUCTIONS}

Table 1 lists the observational details of all 12 SNe recovered in the Snapshot
program GO-8602, while Figure 1 shows the finding charts for the SNe. All the
Snapshot images were obtained in pairs to facilitate removal of cosmic rays.

To properly identify the SNe in the WFPC2 images, we have used the best
available astrometry in the literature (details of the sources for astrometry
can be found in \S~3), and derived the position of the SNe using the pointing
information in the image headers. In most cases, each of the SNe identified in
Figure 1 is the only object within a 0$\farcs 5$ radius error circle.  Three
exceptions are SNe 1997bs, 1999gi, and 1999gq. Identification of SN 1997bs in
the Snapshot images was accomplished by comparing them to the finder chart in
Van Dyk \etal~(2000), where SN 1997bs was apparent in a co-added deep $HST$
archival WFPC2 F555W image.  SN 1997bs is only marginally detected in our F555W
image taken on 2001 March 4 UT (UT dates are used throughout this paper), and
not detected in the two F814W images (see details in \S~3).  Three stars are
within the 0$\farcs 5$ radius error circle of SN 1999gi; following Smartt
\etal~(2001), we identify the brightest object as the SN. No object is within
the 0$\farcs 5$ radius error circle of the position of SN 1999gq.  However, our
estimate for the brightness of SN 1999gq (by comparison with the light curve of
the SN II-P 1988A; Turatto et al. 1993) is 22.0$\pm$0.3 mag at the time of the
Snapshot observation, so the SN should be visible in the image. Fortunately, we
have several excellent KAIT images of SN 1999gq, and after rotating one of them
to the orientation of the WFPC2 image, we measured the relative angular
distance between the SN and several bright stars from the KAIT image, and
identified the SN using the corresponding offsets from these bright stars seen
in the WFPC2 four-chip mosaic. We find that the pointing for the Snapshot
observations of SN 1999gq is off by more than 6$\farcs 0$, relative to
information in the data header.

To measure SN magnitudes in the Snapshot images, we have used the ``HSTphot"
package developed by Dolphin (2000a).  The data quality images are used to mask
bad pixels and defects in the data image using the program {\it mask}. The
program {\it crmask} is used to remove cosmic rays, and the program {\it coadd}
is used to combine the two images. The program {\it getsky} is then used to
determine the sky value at each pixel. Hot pixels are identified and masked
using the program {\it hotpixels}. Following all this preprocessing, the images
are fed into the program {\it hstphot}, where the magnitudes in the WFPC2
system are measured for all the stars using the zero points and charge-transfer
efficiency (CTE) determined by Dolphin (2000b).  We have used option flag 10
for {\it hstphot}, which is a combination of turning on local sky determination
and turning off aperture corrections. Turning on local sky determination is
recommended by the HSTphot manual\footnote{See URL
http://www.noao.edu/staff/dolphin/hstphot/}, while turning off aperture
corrections is necessary because for most of the Snapshot images, there are not
enough bright and isolated stars to determine a reliable aperture
correction. Default aperture corrections are applied to the photometry obtained
with the {\it HST} filters; these are probably accurate, in general, to 0.02
mag in F555W and F814W, and to 0.05 mag in F675W (cf. the HSTphot
manual). These uncertainties are combined in quadrature with the photometric
uncertainties returned by {\it hstphot}, and the resulting uncertainties are
reported in the last column of Table 1.

   Ideally, to tie the magnitudes in the WFPC2 system (Table 1) to the
early-time light curves of the SNe obtained from the ground [which are
typically in the standard Johnson ($UBV$) and Cousins ($RI$) system], a
magnitude transformation such as those described by Holtzman \etal~(1995) and
Dolphin (2000b) should be applied. (Note, however, that although these
transformations are generally applicable to stars, they might not be as
applicable to emission-line dominated sources, such as SNe.) We have done this
for SN 2000cx (see discussion below), but not for the other SNe because of the
difficulty of getting late-time color information for them.  The differences in
these two magnitude systems can be estimated using the procedure described by
Dolphin (2000b) [his equation (1) and Tables 6 and 7]:

$$ V - F555W =  -0.052 (V - I) + 0.027 (V - I)^2,$$ 
$$ I - F814W =  -0.062 (V - I) + 0.025 (V - I)^2,$$

{\noindent where $V$ and $I$ are the magnitudes in the standard
ground-based system, and $F555W$ and $F814W$ are the magnitudes in the
WFPC2 system.  A reasonable range for the late-time $V-I$ color of
the SNe is $-$0.5 to 2 mag, which indicates that the differences
between $V$ and $F555W$, as well as between $I$ and $F814W$, are
smaller than 0.04 mag. In subsequent discussions, we will neglect the
difference between these two systems. }

For SN 2000cx, since the F675W and F814W observations were taken at
the same time, we can use the following equations (also from Dolphin
2000b) to transform the magnitudes:

$$ R - F675W = +0.273 (R - I) -0.066 (R - I)^2, $$
$$ I - F814W = -0.112 (R - I) +0.084 (R - I)^2. $$

The $(R - I)$ color should have minimal difference from its WFPC2
system equivalent $(F675W - F814W)$, which is 0.55 mag from Table
1. When this color is applied to the above equations, we get $R$ =
22.10$\pm$0.06 and $I$ = 21.38$\pm$0.04, which are the magnitudes
for the SN 2000cx Snapshot observations used in subsequent
discussions.

\section{RESULTS}

Figures 2 -- 5 show the light curves of the 12 SNe recovered in the
Snapshot observations. The solid circles are the photometry from
$HST$, while the open circles represent ground-based measurements.
As discussed above, the F555W and $V$-band magnitudes are plotted
together, as are F814W with the $I$ band.  The sources of the ground-based
photometry are listed below in the notes on individual SNe.

\subsection{SN 1995N}

  SN 1995N in MCG $-$02$-$38$-$017 (also Arp 261) was discovered by Pollas
(1995) on 1995 May 5. Optical spectra of the object obtained by Benetti,
Bouchet, \& Schwarz (1995) and by Garnavich \& Challis (1995) showed it to be a
Type IIn SN; specifically, it resembles SN 1988Z in its characteristics (see
Filippenko 1997 for a summary of SNe~IIn). Van Dyk \etal~(1996a) reported the
detection of radio emission from the SN at 3.6 cm on 1995 June 16 and also
provided a precise position of the SN at $\alpha$ = 14$^h$49$^m$28$^s$.313,
$\delta = -10\arcdeg 10\arcmin 13\farcs 92$ (equinox J2000.0), which are the
coordinates we used to locate the SN in the Snapshot observations.

   The ground-based photometry (open circles in Figure 2) consists of
previously unpublished observations taken at the 1.0-m Nickel telescope at Lick
Observatory except the two latest $V$-band points, which were obtained from
Schaefer \& Roscherr (1999) and Schaefer (2001).  In particular, the Schaefer
(2001) observations (which yielded $V$ = 21.1$\pm$0.3 mag) with the McDonald
Observatory 2.1-m telescope were made on the same UT date as our $HST$ Snapshot
observation (which yielded $F555W$ = 21.213$\pm$0.023), and the reported
magnitudes agree with each other very well.  The earliest Nickel observations
taken on 1995 May 23 (18 days after discovery) yielded a $V$ mag of 18.72,
about 1.2 mag fainter than the reported brightness at the time of discovery
(Pollas 1995).  We suspect that the discovery magnitude is erroneous; Fransson
et al. (2002) argue that the SN was already about 10 months old at the time of
discovery, and it declined only by another 1.2 mag in $\sim 1400$ days after
the first Nickel observation.


  As noted by Schaefer \& Roscherr (1999) and Schaefer (2001), SN 1995N has a
very slow optical decline rate, only 2.5 mag (3~from the Nickel photometry) in
the $V$ band over $\sim$2200 days after discovery. This is consistent with the
slow spectral evolution reported by Fransson et al. (2002). The decline rate
for the $V$ band increased from ($0.08 \pm 0.01$ mag)/(100 days) (from
discovery to about 1400 days after discovery) to ($0.18 \pm 0.01$ mag)/(100
days) (from about 1400 days to 2200 days after discovery), while the decline
rate for the $I$ band decreased from ($0.13 \pm 0.01$ mag)/(100 days) (from
discovery to about 1900 days after discovery) to ($0.05 \pm 0.01$ mag)/(100
days) (from about 1900 to 2200 days after discovery). Although the decline
rates in the two passbands seem to change in opposite directions, the light
curves are not sufficiently well sampled to make any definitive conclusions.

Roscherr \& Schaefer (2000) investigated the light echoes from dust formed in
the circumstellar wind of SNe~IIn, and concluded that they cannot properly
account for the slow decline seen in SNe~IIn 1988Z and 1997ab. They suggested
that the shock interaction of the SN ejecta colliding with the circumstellar
wind is the dominant source of late-time emission of these two SNe~IIn,
consistent with previous suggestions (e.g., Filippenko 1997; Aretxaga et
al. 1999).  As we discuss in subsequent sections, the late-time decline rates
for SNe~IIn span a wide range, so there may be differences in the degree of
circumstellar interaction among these SNe.

Fransson \etal~(2002) investigated the late-time spectral evolution of SN 1995N
from both ground-based and $HST$ observations, and concluded that the late-time
emission of SN 1995N is most likely powered by X-rays from the interaction of
the ejecta and the circumstellar medium of the progenitor. They proposed that
the progenitors of SNe~IIn are similar to red supergiants in their superwind
phases, when most of the hydrogen-rich gas is expelled in the last $\sim$
10$^4$ years before explosion.

\subsection{SN 1996cb}

SN 1996cb in NGC 3510 was first discovered by M. Aoki (Nakano \& Aoki 1996) on
1996 December 15, and then independently by BAOSS (Qiao \etal~1996) on 1996
December 18.  Garnavich, Kirshner, \& Berlind (1996a) classified SN 1996cb as a
SN~II near maximum brightness from a spectrum obtained 2 days after discovery;
using a spectrum obtained 16 days later, however, they suggested that it was a
Type IIb SN (Garnavich et al. 1996b), similar to SN 1993J (e.g., Filippenko,
Matheson, \& Ho 1993). An early radio detection was reported by Van Dyk
\etal~(1996b), who also provided a precise position for SN 1996cb as $\alpha$ =
11$^h$3$^m$42$^s$.00, $\delta = +28\arcdeg 54\arcmin14\farcs 2$ (equinox
J2000.0).

Qiu \etal~(1999) reported early spectroscopic and photometric ($BVR$)
observations of SN 1996cb, and concluded that although SN 1996cb is indeed a
SN~IIb similar to SN 1993J, the two SNe have some differences in both the
photometric and spectral evolution. Compared to SN 1993J, SN 1996cb shows
broader light curves, as well as earlier and stronger He~I emission; thus, SN
1996cb may have thicker hydrogen and helium layers than SN 1993J in the outer
ejecta.

The early $V$-band photometry in Figure 2 is taken from Qiu \etal~(1999).
There is no $I$-band photometry for SN 1996cb available in the literature, and
the point shown in Figure 2 is obtained by assuming SN 1996cb has the same
$V-I$ color at the second peak as SN 1993J (Richmond et al. 1996). No reliable
late-time decline rate can be obtained for SN 1996cb, since there is a 1400 day
gap between the two latest $V$-band observations, and there is only a 13-day
difference between the two F814W $HST$ Snapshot observations.  However, it can
be inferred from the latest point in the $V$ band light curve that the decline rate of SN
1996cb does become significantly slower at late times, probably caused by the
interaction between the SN ejecta and the circumstellar medium. Evidence of
circumstellar interaction is also found in the case of SN 1993J, the
prototypical SN~IIb (Filippenko, Matheson, \& Barth 1994; Matheson \etal~2000).

\subsection{SN 1997bs}

SN 1997bs in NGC 3627 was discovered by LOSS (Treffers \etal~1997) on 1997
April 15, the first LOSS SN discovery. A spectrum obtained by Filippenko,
Barth, \& Gilbert (1997) showed that the SN is a peculiar SN~IIn. A precise
position of the SN was provided by Cavagna \& Manca (1997) as $\alpha$ =
11$^h$20$^m$14$^s$.25, $\delta = +12\arcdeg 58\arcmin19\farcs 6$ (equinox
J2000.0).

Van Dyk \etal~(2000) reported early photometry of SN 1997bs from both KAIT and
archival $HST$ WFPC2 observations, some of which are shown in Figure 2. Based
on these observations, Van Dyk \etal~questioned the identification of SN 1997bs
as a bona fide SN, and suggested that it is more likely to be a
``superoutburst" of a luminous blue variable star, analogous to $\eta$ Carinae,
and similar to SN 1961V in NGC 1058 (Goodrich et al. 1989; Filippenko
\etal~1995) and SN 1954J in NGC 2403 (Smith, Humphreys, \& Gehrz 2001). More
recent examples of SN 1997bs-like objects include SN 1999bw (Filippenko, Li, \&
Modjaz 1999; see discussion below), SN 2000ch (Filippenko 2000a; Wagner
\etal~2000), and SN 2001ac (Matheson \& Calkins 2001).  Van Dyk \etal~(2000)
suspected that the progenitor of SN 1997bs might have survived the outburst,
since the SN was seen in early-1998 $HST$ images at $F555W = 23.4$ mag, about
0.5 mag fainter than the progenitor identified by Van Dyk \etal~(1999b) in a
prediscovery image.

The Snapshot observations seem to provide evidence against the superoutburst
interpretation of SN 1997bs. SN 1997bs is marginally detected in the F555W
image taken on 2001 March 4 ($F555W = 25.8 \pm 0.3$ mag), and not detected in
the two F814W images taken on 2001 February 24 and May 28 (limiting magnitude
about 25.0). These observations show that SN 1997bs continued to decline after
early 1998, at 0.21 mag/(100 days) in the $V$ band, and $>$ 0.41 mag/(100 days)
in the $I$ band, inconsistent with the suggestion that the progenitor of the SN
survived the explosion.  While formation of dust in the ejecta could explain
the continuous decline in the optical wavelengths ($V$ and $I$), it is
inconsistent with the color evolution. The SN should become progressively
redder if dust is forming in the ejecta, but SN 1997bs has become progressively
bluer, from $V - I = 3.0$ mag in early 1998 to $V - I < 0.8$ mag in early 2001.
Additional deep multiband $HST$ images of SN 1997bs could provide useful
information on the late-time behavior of the object and help to constrain its
nature.

\subsection{SN 1997ef}

SN 1997ef in UGC 4107 was discovered by Y. Sano (Nakano \& Sano 1997) on 1997
November 25. Garnavich \etal~(1997a) obtained a spectrum of the object that
showed very unusual broad features, and failed to identify it as any known type
of SN. Filippenko \& Martin (1997) suggested that the object may be a
previously unobserved, extreme example of a stripped (Type Ic-like) SN; this
was also proposed by Garnavich \etal~(1997b) and by Wang, Howell, \& Wheeler
(1998).  A precise position of the SN was provided by Nakano \& Sano (1997):
$\alpha$ = 7$^h$57$^m$2$^s$.82, $\delta = +49\arcdeg 33\arcmin40\farcs 2$
(equinox J2000.0).

SN 1997ef is the subject of a number of studies, because of its peculiar
spectroscopic behavior and its possible association with a gamma-ray burst that
occurred on 1997 November 15 (e.g., Wang \& Wheeler 1998; Nomoto \etal~1999;
Iwamoto \etal~2000; Mazalli, Iwamoto, \& Nomoto 2000). Iwamoto \etal~(2000)
also claim that SN 1997ef belongs to a class of objects termed ``hypernova,"
the prototype of which is SN 1998bw (e.g., Iwamoto \etal~1998; Woosley,
Eastman, \& Schmidt 1999).

The early $I$-band photometry of SN 1997ef shown in Figure 3 consists of
previously unpublished KAIT observations. These data show that SN 1997ef
declined at a rate of 1.27 mag/(100 days) in the first 100 days after
discovery. The decline rate changed dramatically thereafter, and the SN faded
only 2.5 mag over the next 1100 days. Again, circumstellar interaction or a
light echo could be a possible cause.

\subsection{SN 1998S}

SN 1998S in NGC 3877 was discovered by BAOSS on 1998 March 3 (Li \& Li 1998).
A spectrum taken by Filippenko \& Moran (1998) indicated that SN 1998S was a
peculiar SN~IIn. Van Dyk \etal~(1999c) reported radio detection of SN 1998S on
1999 October 28, and provided a precise position as $\alpha$ =
11$^h$46$^m$6$^s$.140, $\delta = +47\arcdeg 28\arcmin55\farcs 45$ (equinox
J2000.0).

The early-time photometry of SN 1998S shown in Figure 3 consists of KAIT
observations taken from Modjaz \etal~(2002, in preparation).  Early optical and
infrared photometry of SN 1998S was also reported by Fassia \etal~(2000) and
Liu \etal~(2000). The
$V$-band and $I$-band decline rates for SN 1998S seem to exhibit two
significant changes, the first at $\sim$100 days after maximum brightness, and
the second roughly 300 days after maximum. The late-time ($>$ 300 d) decline
rate of SN 1998S is ($0.34 \pm 0.03$ mag)/(100 days) in the $V$ band, and
($0.56 \pm 0.05$ mag)/(100 days) in the $I$ band. This slow decline rate is
certainly caused by strong circumstellar interaction, as discussed by Leonard
\etal~(2000) based on spectropolarimetry and spectroscopy of SN 1998S, as well
as from its radio and X-ray emission (Pooley \etal~2002).

\subsection{SN 1999bw}

SN 1999bw in NGC 3198 was discovered by LOSS (Li 1999). A precise position was
also provided by Li (1999) as $\alpha$ = 10$^h$19$^m$46$^s$.81, $\delta =
+45\arcdeg 31\arcmin 35\farcs 0$ (equinox J2000.0). Spectra taken by Garnavich
\etal~(1999a) and Filippenko \etal~(1999) showed barely resolved hydrogen
Balmer emission lines, and Filippenko, Li, \& Modjaz (1999) classified the
object as a SN 1997bs-like Type IIn SN.

The early-time photometry of SN 1999bw shown in Figure 3 consists of previously
unpublished KAIT observations. SN 1999bw declined 5.6 mag in the first 620 days
after maximum, while SN 1997bs declined about 7.0 mag in the same period
(estimated from Figure 2), so their photometric evolution seems to differ to
some extent, even though they have similar spectra and low peak luminosity
(Filippenko, Li, \& Modjaz 1999).

\subsection{SN 1999eb}

SN 1999eb in NGC 664 was discovered by LOSS (Modjaz \& Li 1999) on 1999 October
2.  The SN was also present in earlier LOSS images taken on 1999 September 22
and 29.  A precise position is reported by Modjaz \& Li (1999) as $\alpha$ =
1$^h$43$^m$45$^s$.45, $\delta = +4\arcdeg 13\arcmin25\farcs 9$ (equinox
J2000.0).  Garnavich \etal~(1999b) obtained a spectrum of the SN and classified
it as a Type IIn. Terlevich, Fabian, \& Turatto (1999) suggested that SN 1999eb
was possibly associated with a gamma-ray burst that occurred on 1999 October 2.
However, as pointed out by Filippenko (2000b), this association seems quite
unlikely, given the fact that SN 1999eb was present in LOSS images on 1999
September 22, 10 days before the gamma-ray burst.

The early-time photometry of SN 1999eb shown in Figure 4 consists of KAIT data
taken from Modjaz \etal~(2002, in preparation). The F555W observation $\sim$450
days after maximum brightness is in good agreement with a linear extrapolation
from early points, suggesting that the decline rate in the $V$ band probably
does not change significantly during this period.  There are three $F814$
observations for SN 1999eb, which yield a good estimate of the late-time
decline rate as ($0.86 \pm 0.01$ mag)/(100 days).

\subsection{SN 1999el}

SN 1999el in NGC 6951 was discovered by BAOSS (Cao \etal~1999) on 1999 October
20.  A precise position was also provided by Cao \etal~(1999) as $\alpha$ =
20$^h$37$^m$17$^s$.83, $\delta = +66\arcdeg 6\arcmin 11\farcs 5$ (equinox
J2000.0).  From a spectrum taken with the BAO 2.16-m telescope, Filippenko
(1999) identified SN 1999el as a very early Type IIn SN.

The early-time photometry of SN 1999el shown in Figure 4 consists of KAIT data
taken from Modjaz \etal~(2002, in preparation). The late-time Snapshot
observations suggest that the decline rate of SN 1999el in both the $V$ and $I$
bands has changed at late times. The two F814W Snapshot observations yield a
late-time $I$-band decline rate of ($1.30 \pm 0.03$ mag)/(100 days).

\subsection{SN 1999gi}

SN 1999gi in NGC 3184 was discovered by R. Kushida (Nakano \& Kushida 1999) on
1999 December 9. A precise position is also reported by Nakano \& Kushida
(1999) as $\alpha$ = 10$^h$18$^m$16$^s$.66, $\delta = +41\arcdeg 26\arcmin
28\farcs 2$ (equinox J2000.0). Jha \etal~(1999a) obtained a spectrum of SN
1999gi and classified it as a SN~II.

The well-sampled early-time light curves for SN 1999gi shown in Figure 4 are
the KAIT observations taken from Leonard \etal~(2002b), and indicate that SN
1999gi is a normal Type II-P SN. The F555W observation is consistent with a
linear extrapolation of the latest KAIT data, and yields a decline rate of
($1.01 \pm 0.01$ mag)/(100 days).

Our Snapshot image of SN 1999gi was used by Smartt \etal~(2001) to place an
upper limit on the mass of the progenitor of the SN. However, as discussed more
fully by Leonard \etal~(2002b), their derived upper limit may be too stringent,
since the distance they used to the SN (7.9 Mpc) is substantially less than the
distance derived through the expanding photosphere method ($\sim 12$ Mpc) and
other recent distance estimates to the host galaxy.

\subsection{SN 1999gq}

SN 1999gq in NGC 4523 was discovered by LOSS (Papenkova \& Li 1999) on 1999
December 23. An independent discovery was reported by Armstrong (1999).  SN
1999gq was classified as a SN~II from spectroscopic observations by Ayani \&
Yamaoka (1999) and by Jha \etal~(1999b). A precise position of the SN was
reported by Papenkova \& Li (1999) as $\alpha$ = 12$^h$33$^m$48$^s$.32, $\delta
= +15\arcdeg 10\arcmin48\farcs 2$ (equinox J2000.0).

The sparse early-time photometry of SN 1999gq shown in Figure 5 consists of
previously unpublished KAIT observations. The SN has not been well observed,
and the gap between the early-time and Snapshot observations is too
large ($>$ 500 days) to determine a reliable decline rate.

\subsection{SN 2000P}

SN 2000P in NGC 4965 was discovered by R. Chassagne (Colas \& Chassagne 2000)
on 2000 March 8. A precise position of the SN is reported by Colas \& Chassagne
(2000) as $\alpha$ = 13$^h$7$^m$10$^s$.53, $\delta = -28\arcdeg
14\arcmin2\farcs 5$ (equinox J2000.0). SN 2000P was spectroscopically
classified as a SN~IIn (Cappellaro \etal~2000; Jha \etal~2000).

The early-time photometry of SN 2000P shown in Figure 5 consists of previously
unpublished KAIT observations. The F555W observation 250 days after discovery
indicates that the decline rate of SN 2000P changed at late times.

\subsection{SN 2000cx}

SN 2000cx in NGC 524 was discovered by LOSS (Yu, Modjaz, \& Li 2000)
on 2000 July 17. Yu et al. also provided a precise position of
the SN as $\alpha$ = 1$^h$24$^m$46$^s$.15, $\delta = +09\arcdeg
30\arcmin30\farcs 9$ (equinox J2000.0).  SN 2000cx was classified as a
SN 1991T-like peculiar SN~Ia from a spectrum obtained by Chornock
\etal~(2000).

SN 2000cx is one of the best-monitored SNe~Ia in the literature. The early
spectroscopic and photometric observations were reported by Li \etal~(2001a),
which showed SN 2000cx to be a unique SN~Ia. SN 2000cx showed an apparent
asymmetry in the $B$-band light curve, in which the premaximum brightening is
relatively fast (similar to that of the normal SN 1994D), but the postmaximum
decline is relatively slow (similar to that of the overluminous SN 1991T). The
color of SN 2000cx is also extremely blue. The premaximum spectra of SN 2000cx
are similar to those of SN 1991T-like objects, but its overall spectral
evolution is quite different, with strong Si~II lines until three weeks past
$B$ maximum and a slow change in the excitation stages of iron-peak
elements. Matter is also moving at high expansion velocities in the ejecta of
SN 2000cx. Based on these observations, Li \etal~proposed that SN 2000cx may be
an overluminous object like SN 1991T, but with a larger yield of $^{56}$Ni and
a higher kinetic energy in the ejecta.

The early-time photometry of SN 2000cx in Figure 5 consists of the KAIT
observations from Li \etal~(2001a). The Snapshot observations of SN 2000cx were
made about a year after discovery, which show that the decline rate of the SN
in both the $R$ and $I$ bands has decreased at late times. The decline rate
measured from the latest KAIT points and the Snapshot observations is ($1.61
\pm 0.05$ mag)/(100 days) for the $R$ band, and ($1.10 \pm 0.08$ mag)/(100
days) for the $I$ band.  The change of decline rate at late times for SN 2000cx
is most likely not due to light echoes, since SN 2000cx occurred at the
outskirts of an early-type (S0) galaxy, where a dusty environment is not
expected. Light echoes have been successfully detected around at least two
SNe~Ia: SN 1991T, after 600 days following maximum (Schmidt \etal~1994), and SN
1998bu, after 500 days following maximum (Cappellaro \etal~2001).

\subsection{Comparison of SNe~IIn}

Seven of the 12 SNe recovered in our Snapshot program are SNe~IIn and show slow
evolution at late times. As discussed in previous sections, some differences
among their late-time decline rates are found.  This is to be expected because
the subclass of SNe~IIn is known to span a very broad range of properties
(e.g., Filippenko 1997; Filippenko, Li, \& Modjaz 1999). Some SNe~IIn are
subluminous (e.g., SN 1997bs with $M_V = -13.8 $ mag, Van Dyk \etal~2000),
others are moderately bright (e.g., SN 1999eb with $M_V = -18.6 $ mag, assuming
$H_0$ = 65 km s$^{-1}$ Mpc$^{-1}$), while a few can rival the brightness of a
SN~Ia (e.g., SN 1998S with $M_V = -19.6 $ mag, Leonard \etal~2000).

Figure 6 shows the light curves of all the SNe~IIn recovered in our Snapshot
program. Even though the data are often limited by a large gap in time between
the ground-based and the $HST$ Snapshot photometry, or by large time intervals
between the Snapshot observations, it is still easily seen from Figure 6 that
the light-curve shapes for the SNe~IIn are quite different: some decline
monotonically after maximum (e.g., SN 1999eb in the $V$ band, and SN 1999eb in
the $I$ band), while others experience multiple changes in their decline rate
(e.g., SNe 1997bs and 1998S in the $V$ band).  There is no clear correlation
between the light-curve shapes and the luminosities of the SNe~IIn: the
overluminous SN 1998S and the subluminous SN 1997bs have similar narrow peaks,
while the moderately bright SN 1999eb has a very broad peak.

Table 2 lists the late-time decline rates for some of the SNe~IIn, and it is
clear that they span quite a large range.  Note, however, that these are
sometimes derived at different evolutionary phases of the SNe, so a close
comparison should not be made.

We suspect that the different light-curve shapes of SNe~IIn are closely related
to their progenitor properties. Relevant factors may be the density, structure,
and radial distance of the circumstellar medium, the mass and composition of
the progenitor, and so forth.

\subsection{Contamination of SN Light in Template Images}

The two vertical dotted lines in Figure 6 mark one and two years after maximum
light. It can be seen that the SNe declined 0.3 to 7.3 mag in the $V$ band, and
0.3 to 6.2 mag in the $I$ band at one year after discovery, while they declined
0.6 to 8.5 mag in the $V$ band, and 0.8 to 8.3 mag in the $I$ band at two years
after discovery. These numbers indicate that the template images for these
SNe~IIn taken at one and two years after discovery will have different degrees
of contamination from the SN light. For the extreme case of SN 1995N, which had
declined only 0.6 mag in the $V$ band and 0.8 mag in the $I$ band by two years
after discovery, the template images are seriously contaminated by the
SN. While SN 1995N is at a fortunate location without a complicated background
and the point-spread-function technique used to derive the photometry reported
in \S~3.1 is adequate, other SNe having similar behavior may not be so
fortunate and it will take a very long time to obtain template images without
significant contamination from the SN light. On the other hand, SN 1998S
declined 7.3 mag in the $V$ band and 6.2 mag in the $I$ band by one year after
discovery, which means the SN was only 0.30\% and 0.33\% of the peak $V$ and
$I$-band brightness, respectively; thus, the images taken one year after
discovery will be good templates for those taken around maximum light.

We have well-defined late-time light curves for a few other SNe in our
sample. For SN Ia 2000cx, we measured a decline of 9.0 mag in the $R$ band and
8.1 mag in the $I$ band by one year after discovery, which indicates that the
late-time templates are not seriously contaminated by the SN light.  For SN
II-P 1999gi, we measured a decline of 4.6 mag in the $V$ band by one year after
discovery, when the SN was about 1.5\% of its peak brightness. High-precision
photometry of SNe II-P thus requires the template images to be taken at a time
considerably later than one year after discovery.

\section{CONCLUSIONS}

In this paper we present 12 SNe recovered in the $HST$ Cycle 9 Snapshot program
GO-8602. $HST$ photometry was obtained for the SNe, and the data were tied to
early-time ground-based observations, in order to study the late-time evolution
of various types of SNe. Much of the ground-based data presented here are previously
unpublished, and were obtained primarily with the Katzman Automatic Imaging
Telescope.

The successful detection of SN 1996cb, a SN~IIb, more than 4 years after
discovery suggests the existence of interaction between the SN ejecta and the
circumstellar medium.  The peculiar SN~Ic 1997ef, a ``hypernova," is found to
decline slowly between 100 and 1200 days after discovery.  Two bright SN II-P
discovered in 1999, SNe 1999gi and 1999gq, are also both recovered in the
Snapshot observations.

We find that there is a diversity in the late-time evolution of SNe~IIn,
perhaps due to differences in circumstellar interaction. Some decline
monotonically, such as SN 1999eb within the first 660 days following maximum
brightness. Others experience multiple changes in their decline rate, such as
SNe 1998S, 1999el, and 2000P. SN 1995N fades very slowly, declining by only 1.2
mag in $V$ over $\sim 1400$ days following discovery. Thus, template images of
some SNe must be obtained many years after the explosion, to minimize
contamination from the SN itself.

 SN 1997bs, a subluminous SN~IIn that was suggested by Van Dyk \etal (2000) to
be a superoutburst of a massive luminous blue variable, is only marginally
detected in the F555W observations and not at all in the two F814W images,
casting some doubt on the hypothesis that the progenitor survived the
explosion.  SN 1999bw, an object similar to SN 1997bs in terms of its spectrum
and peak luminosity, showed a slower late-time decline rate than SN 1997bs.

  SN 2000cx is the only SN~Ia recovered in the Snapshot observations.  Both the
$R$-band and $I$-band decline rates are found to decrease at late times, but we
do not believe this is due to a light echo.

  Circumstellar interaction is apparently very common in the late-time
evolution of core-collapse SNe, such as SNe~IIn, SNe~IIb, and possibly some
SNe~II-P and SNe~Ic.

\acknowledgments

We thank the staff of the Lick Observatory for their assistance, and we
acknowledge useful conversations with M. W. Richmond and R. Chornock.  We also
thank P. Garnavich, A. Pastorello, M. Turatto, and B. Schaefer for providing us
with their calibrations of the SN 1995N field. Support for proposal number
GO-8602 was provided by NASA through a grant from the Space Telescope Science
Institute, which is operated by AURA, Inc., under NASA contract NAS 5-26555;
additional funding was provided by NASA grants GO-8648 and GO-9114.  The work
of A.V.F.'s group at the University of California, Berkeley is also supported
by National Science Foundation grant AST-9987438, as well as by the Sylvia and
Jim Katzman Foundation. KAIT was made possible by generous donations from Sun
Microsystems, Inc., the Hewlett-Packard Company, AutoScope Corporation, Lick
Observatory, the National Science Foundation, the University of California, and
the Katzman Foundation.  A.V.F. is grateful to the Guggenheim Foundation for a
Fellowship.

\newpage

\begin{figure} [b!]
{\plotfiddle{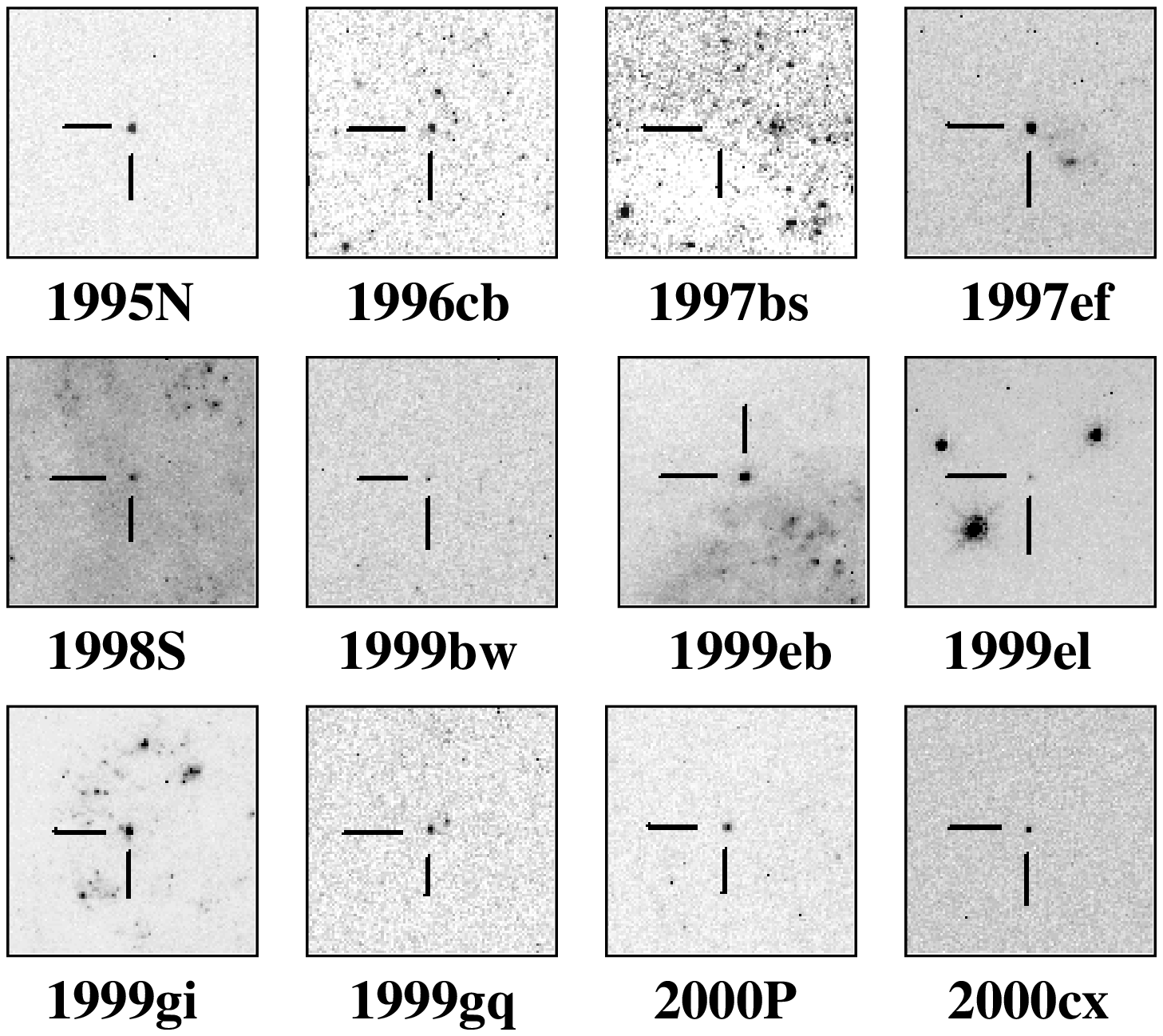}{8.2in}{0}{100}{100}{-30}{-60}}
\caption{Finder charts for the 12 SNe recovered in the Snapshot
observations. Each panel consists of 156$\times$156 PC pixels, and is
$7\farcs 2\times7\farcs 2$ in size.}  
\label{1}
\end{figure} 

\newpage

\begin{figure}
{\plotfiddle{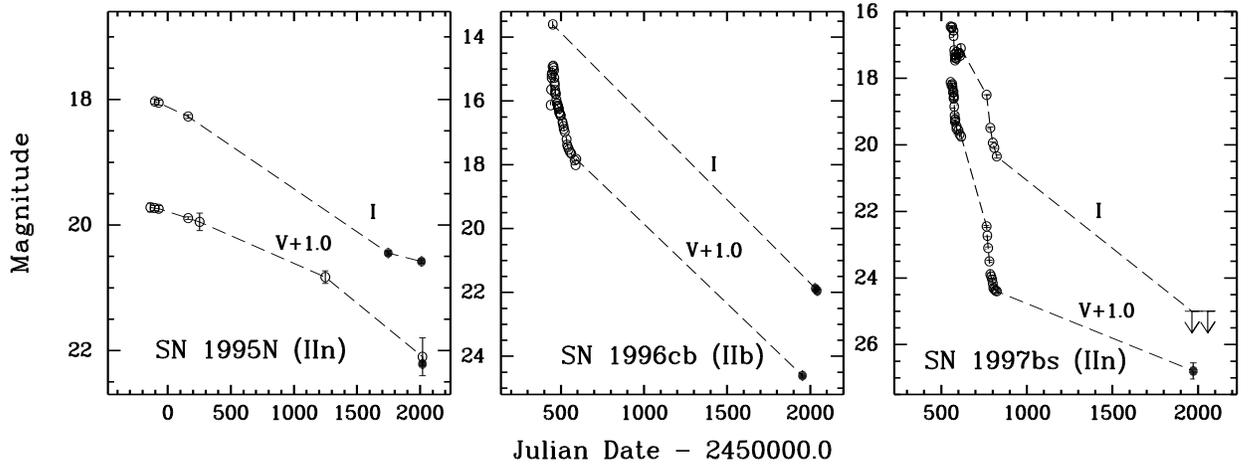}{6.2in}{-90}{65}{65}{0}{390}}
\caption{The light curves of SNe 1995N, 1996cb, and 1997bs. The open circles
are the photometry from ground-based observations (see text for sources), and
the solid circles are data obtained from the $HST$ Snapshot observations. An
offset has been added to the $V$-band data for clarity. Dashed lines between
widely separated points are drawn to help guide the eye; they represent only
the average decline rates, not the detailed light curves. }
\label{2}
\end{figure}

\begin{figure}
{\plotfiddle{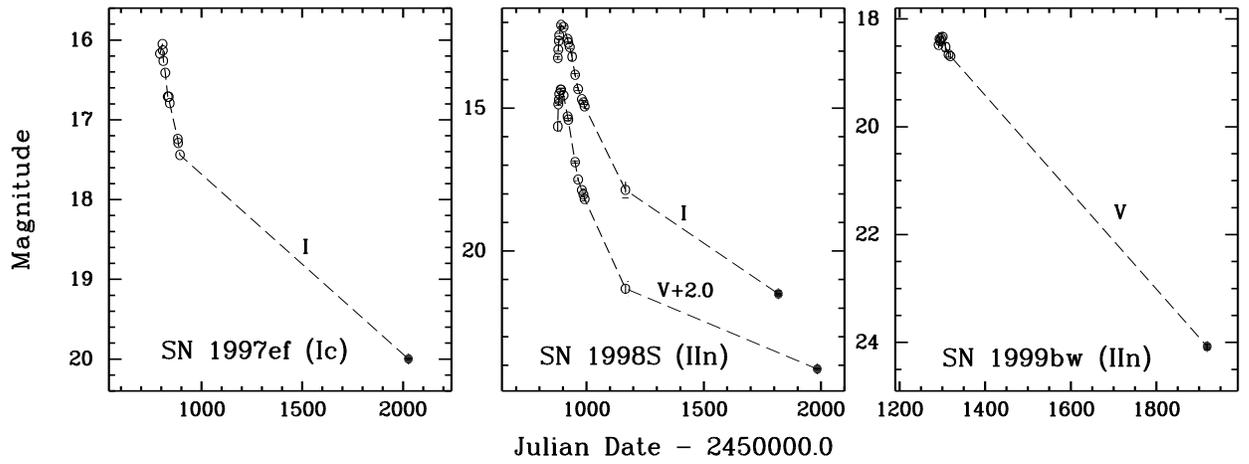}{6.2in}{-90}{65}{65}{0}{390}}
\caption{The same as Figure 2, but for SNe 1997ef, 1998S, and 1999bw.}
\label{3}
\end{figure}

\begin{figure}
{\plotfiddle{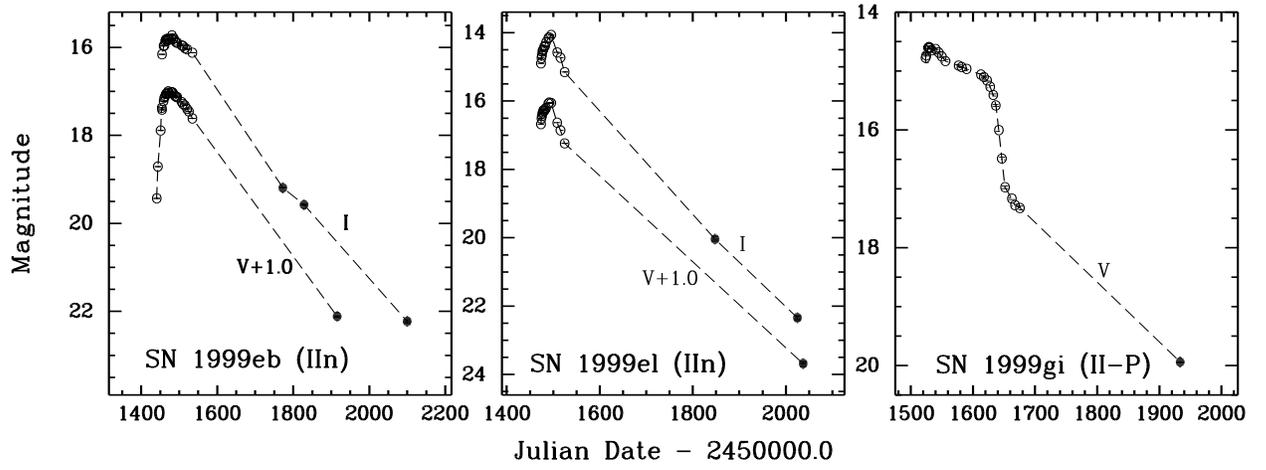}{6.2in}{-90}{65}{65}{0}{390}}
\caption{The same as Figure 2, but for SNe 1999eb, 1999el, and 1999gi.}
\label{4}
\end{figure}

\begin{figure}
{\plotfiddle{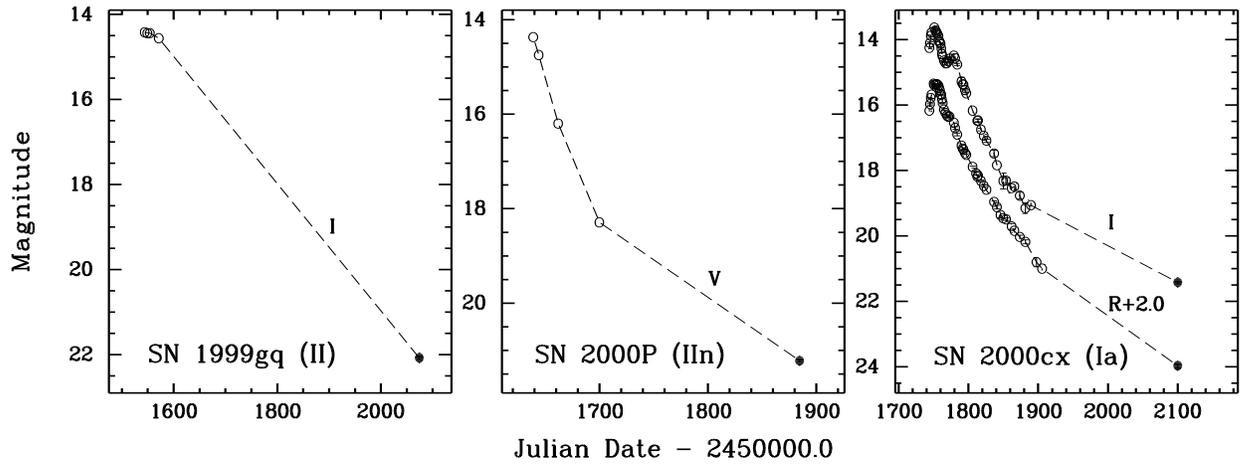}{6.2in}{-90}{65}{65}{0}{390}}
\caption{The same as Figure 2, but for SNe 1999gq, 2000P, and 2000cx.}
\label{5}
\end{figure}

\begin{figure}
{\plotfiddle{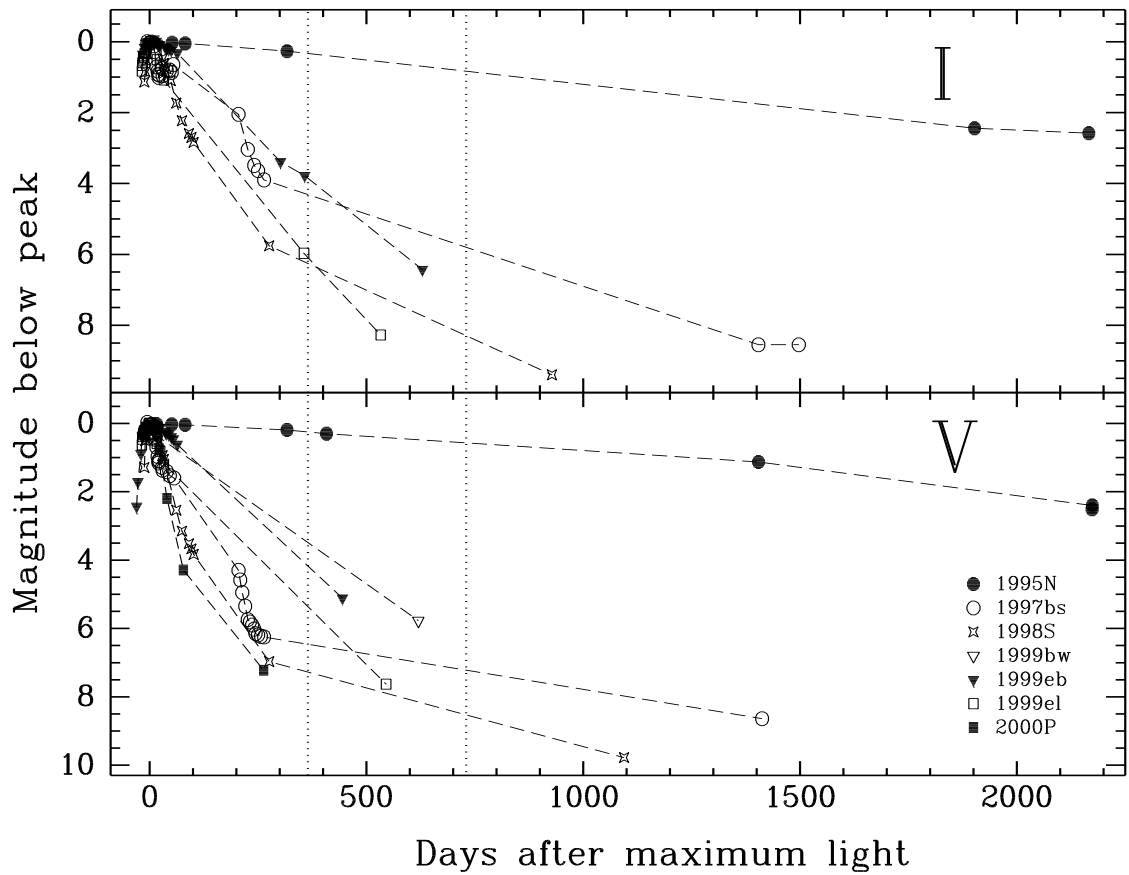}{6.2in}{-90}{65}{65}{0}{390}}
\caption{Comparison of the light curves of SNe~IIn recovered in the 
Snapshot observations. The light curves have been shifted to match 
the time of maximum light and the peak brightness. The two vertical 
dotted lines indicate one and two years after maximum brightness.  
One can see the large variety of photometric evolution among SNe~IIn.}
\label{6}
\end{figure}

\newpage

\renewcommand{\arraystretch}{0.55}

\begin{deluxetable}{llllllll}
\tablecaption{Observational details of the Snapshot SNe}
\tablehead{
\colhead{SN} & \colhead{Type} & \colhead{Date\tablenotemark{a}} & Age\tablenotemark{b} &
\colhead{Filter} & \colhead{Exp\tablenotemark{c}} & \colhead{J.D.\tablenotemark{d}} &
\colhead{Magnitude\tablenotemark{e}} 
}
\startdata
  1995N &IIn&Jul 22, 2000&1903&F814W&700&51747.44&20.45$\pm$0.02 \\
  1995N &IIn&Apr 11, 2001&2166&F814W&700&52010.88&20.58$\pm$0.02 \\
  1995N &IIn&Apr 19, 2001&2174&F555W&700&52018.45&21.21$\pm$0.02 \\
  1996cb&IIb&Feb 13, 2001&1515&F555W&700&51953.18&23.61$\pm$0.07\\
  1996cb&IIb&Apr 30, 2001&1591&F814W&700&52029.93&21.88$\pm$0.05 \\
  1996cb&IIb&May 13, 2001&1604&F814W&700&52042.65&21.95$\pm$0.04 \\
  1997bs&IIn&Mar  4, 2001&1417&F555W&700&51972.46&25.79$\pm$0.25 \\
  1997ef&Ic&Apr 28, 2001&1233&F814W&700&52027.50&20.00$\pm$0.03 \\
  1998S &IIn&Oct  1, 2000&940&F814W&700&51818.03&21.50$\pm$0.03 \\
  1998S &IIn&Mar 16, 2001&1106&F555W&700&51984.16&22.13$\pm$0.03 \\
  1999bw&IIn&Jan  9, 2001&626&F555W&700&51918.58&24.08$\pm$0.06 \\
  1999eb&IIn&Aug 16, 2000&332&F814W&700&51772.68&19.19$\pm$0.03 \\
  1999eb&IIn&Oct 11, 2000&387&F814W&700&51828.42&19.58$\pm$0.03 \\
  1999eb&IIn&Jan  6, 2001&475&F555W&700&51915.61&21.11$\pm$0.03 \\
  1999eb&IIn&Jul  9, 2001&659&F814W&700&52099.94&22.23$\pm$0.04 \\
  1999el&IIn&Oct 30, 2000&374&F814W&700&51847.74&20.04$\pm$0.03 \\
  1999el&IIn&Apr 25, 2001&551&F814W&700&52024.74&22.34$\pm$0.04 \\
  1999el&IIn&May  8, 2001&563&F555W&700&52037.13&22.68$\pm$0.03 \\
  1999gi&II-P&Jan 24, 2001&409&F555W&700&51933.43&19.94$\pm$0.02 \\
  1999gq&II&Jan 14, 2001&530&F814W&700&52074.24&22.07$\pm$0.05 \\
  2000P &IIn&Dec  6, 2000&245&F555W&700&51884.56&21.22$\pm$0.02 \\
  2000P &IIn&Apr 23, 2001&383&F814W&700&52022.60&21.33$\pm$0.03 \\
  2000cx&Ia&Jul 10, 2001&356&F675W&280&52100.07&21.97$\pm$0.06\tablenotemark{f} \\
  2000cx&Ia&Jul 10, 2001&356&F814W&280&52100.07&21.42$\pm$0.04\tablenotemark{g} \\
\enddata
\tablenotetext{a} {UT date of the observation.}
\tablenotetext{b} {Approximate age of the SN in days since discovery.}
\tablenotetext{c} {Exposure time in seconds.}
\tablenotetext{d} {Modified Julian Date $-$ 2,400,000.} 
\tablenotetext{e} {Magnitude in the WFPC2 system. An uncertainty of 0.02
mag is added in quadrature to the errors of the F555W and F814W
observations, while 0.05 mag is added to the F675W error
for SN 2000cx. These magnitudes are converted to Johnson $V$ and Cousins
$I$, except in the case of SN 2000cx.}
\tablenotetext{f} {The transformed standard Johnson $R$ magnitude 
is 22.10$\pm$0.06.}
\tablenotetext{g} {The transformed standard Cousins $I$ magnitude 
is 21.38$\pm$0.04.}
\end{deluxetable} 

\newpage

\begin{deluxetable}{llll}
\tablecaption{Late-time decline rate of several SNe IIn.}
\tablehead{
\colhead{SN} & \colhead{Filter} & \colhead{Decline rate\tablenotemark{a}} &
\colhead{Age\tablenotemark{b}~ (days)}
}
\startdata
1995N &$V$&0.18$\pm$0.01&1400 $-$ 2200 \\
1997bs&$V$&0.21$\pm$0.02&300 $-$ 1400 \\
1998S &$V$&0.34$\pm$0.03&300 $-$ 1100 \\
\tableline
1995N &$I$&0.05$\pm$0.01&1900 $-$ 2200 \\
1998S &$I$&0.56$\pm$0.05&300 $-$ 1000 \\
1999eb&$I$&0.86$\pm$0.01&300 $-$ 600 \\
1999el&$I$&1.30$\pm$0.03&400 $-$ 600 \\
\enddata
\tablenotetext{a} {Magnitude decline per 100 days.}
\tablenotetext{b} {Approximate age of the SN when the decline rate is derived.}
\end{deluxetable} 

\end{document}